 \documentclass[%
 aip,
 amsmath,amssymb,
 reprint,%
]{revtex4-2}

\usepackage[utf8]{inputenc}
\usepackage[T1]{fontenc}
\usepackage{mathptmx}
\usepackage{etoolbox}

\usepackage{graphicx,epsfig}
\usepackage{amsfonts}
\usepackage{xfrac}
\usepackage{xcolor}
\usepackage{epstopdf}
\usepackage[hidelinks]{hyperref}
\usepackage[capitalise]{cleveref}
\usepackage{bbold}
\usepackage{caption}
\usepackage{fancyhdr}
\usepackage{subcaption}
\usepackage{bm}

\newcommand{\Da}{\mathcal{D}}

\newcommand{\Ua}{\mathcal{U}}

\makeatletter
\def\@email#1#2{%
 \endgroup
 \patchcmd{\titleblock@produce}
  {\frontmatter@RRAPformat}
  {\frontmatter@RRAPformat{\produce@RRAP{*#1\href{mailto:#2}{#2}}}\frontmatter@RRAPformat}
  {}{}
}%
\makeatother

\begin{document}

\preprint{AIP/123-QED}

\title{Microscopic field theory for structure formation in systems of self-propelled particles with generic torques}
\author{Elena Ses\'e-Sansa}
\affiliation{CECAM,  Centre  Europ\'een  de  Calcul  Atomique  et  Mol\'eculaire, \'Ecole  Polytechnique  F\'ed\'erale  de  Lausanne (EPFL),  Batochime,  Avenue  Forel  2,  1015  Lausanne,  Switzerland}
 \email{elena.sesesansa@epfl.ch}
\author{Demian Levis}%
\affiliation{ Departament de F\'isica de la Mat\`eria Condensada, Universitat de Barcelona, Mart\'i i Franqu\`es 1, 08028 Barcelona, Spain}%
\affiliation{UBICS  University  of  Barcelona  Institute  of  Complex  Systems,  Mart\'{\i}  i  Franqu\`es  1,  08028  Barcelona,  Spain}
\author{Ignacio Pagonabarraga}
\affiliation{CECAM,  Centre  Europ\'een  de  Calcul  Atomique  et  Mol\'eculaire, \'Ecole  Polytechnique  F\'ed\'erale  de  Lausanne (EPFL),  Batochime,  Avenue  Forel  2,  1015  Lausanne,  Switzerland}
\affiliation{ Departament de F\'isica de la Mat\`eria Condensada, Universitat de Barcelona, Mart\'i i Franqu\`es 1, 08028 Barcelona, Spain}%
\affiliation{UBICS  University  of  Barcelona  Institute  of  Complex  Systems,  Mart\'{\i}  i  Franqu\`es  1,  08028  Barcelona,  Spain}

\date{\today}

\begin{abstract}
We derive a dynamical field theory for self-propelled particles subjected to generic torques and forces by explicitly coarse-graining their microscopic dynamics, described by a many-body Fokker-Planck equation. 
The model includes both intrinsic torques inducing self-rotation, as well as interparticle torques leading to, for instance, the local alignment of  particles' orientations. Within this approach, although the functional form of the pairwise interactions does not need to be specified, one can directly  map the parameters of the field theory onto the parameters of particle-based models. 
We perform a linear stability analysis of the  homogeneous solution of the field equations and find both long-wavelength and short-wavelength instabilities. The former signals the emergence of a macroscopic structure, which we associate with motility-induced phase separation, while the second one
 signals the growth of a finite structure with a characteristic size. Intrinsic torques hinder phase separation, pushing the onset of the long-wavelength instability to higher activities. Further, they generate finite-sized structures with a characteristic size proportional to both the self-propulsion velocity and the inverse of the self-rotation frequency. Our results show that a general mechanism might explain why chirality tends to suppress motility-induced phase separation but instead promotes the formation of non-equilibrium patterns. 
 \end{abstract}

\maketitle


\section{Introduction}


Active matter encompasses a large number of systems which constantly convert internal energy into autonomous motion. 
They thus evolve far from equilibrium, such that the understanding of their collective states resulting from the combination of different interactions between their constituents, in the presence of dissipation and fluctuations violating the standard rules of equilibrium dynamics, still raises interesting challenges. 
Lifting constraints imposed by equilibrium,  such as the fluctuation-dissipation theorem, active matter can reach novel  spatio-temporal structures absent in passive matter \cite{Ramaswamy2010,Bechinger2016}.
Living organisms constitue obvious examples of active matter, although artificially self-propelled objects, such as granular 
 \cite{deseigne2010,briand2018,scholz2018rotating,arora2021emergent} or colloidal 
  \cite{sanchez2015, Buttinoni2012, Palazzi2013, ginot2015nonequilibrium,VanderLinden2019} particles, have become popular in soft matter physics labs over the last decade.

In order to somehow grasp the physics of collections of active particles, quite some efforts have been devoted to the study of simplified models \cite{marchettiRev,WinklerRev}.  In this context, much progress has been achieved in the fundamental understanding of two widespread collective phenomena in active particle systems: on the one hand, the emergence of collective motion, or flocking, and on the other hand, particle clustering in the absence of attractive interactions. 

 The Vicsek model was the first attempt to describe the former phenomenon, i.e. the emergence of collective motion, as a symmetry breaking phase transition, due to the competition between noise and local alignment of the particles' (birds) self-propulsion direction \cite{Vicsek1995, GinelliRev,Vicsek2012}. Its continuum counterpart, the Toner-Tu theory, provides a hydrodynamic description incorporating the same main fundamental ingredients as the agent-based model  \cite{Toner1995,Toner1998}. Such continuum approach allows for an understanding of the generic mechanisms controlling the large-scale properties of systems of aligning self-propelled particles, hoping them to be, to some extent at least, universal. Since then, further efforts have been put into deriving continuum equations in a consistent way, by explicitly coarse-graining the stochastic dynamics of  self-propelled particles \cite{bertin2009, ihle2011kinetic, peshkov2014boltzmann,dadam}. One of the advantages of constructing a field theory starting from the microscopic dynamics is that it allows for a direct comparison between the continuum and particle-based results, something which is missing in the original Toner-Tu approach, based on symmetry and conservation laws assumptions.
Besides the Vicsek model (and its variants), which prescribes a velocity-aligning torque between otherwise non-interacting agents, flocking behavior  can also arise from excluded-volume interactions between elongated self-propelled particles \cite{Peruani2006, abkenar2013,Jayaram2020, Grossmann2020,PeruaniRev} or from other more complex mechanisms involving the coupling with the environment, or the specificities of the particles' self-propulsion mechanism \cite{deseigne2010,bricard2013emergence, yan2016reconfiguring, kaiser2017flocking,chardac2021topology}. 

The other salient phenomenon generically encountered in active systems is the spontaneous aggregation in the absence of attractive interactions \cite{Buttinoni2012,Palazzi2013,Buttinoni2013}. At the level of simple models, this phenomenon is thought to arise from the mere competition between  persistent motion along a given direction and excluded-volume interactions. For persistent enough particles and at high enough densities, one can eventually observe the system phase separate into a macroscopic dense cluster surrounded by a dilute gas-like phase, a phenomenon  known as motility-induced phase separation (MIPS) \cite{Tailleur2008,Cates2015}. MIPS has been reported in numerical studies of minimal models, like  Active Brownian Particles (ABP), consisting in persistent Brownian  spherical  particles, interacting  with, typically, an isotropic short-range repulsive potential \cite{Fily2012,Redner2013,stenhammar2014,Fily2014,Cates2015,Levis2017,Digregorio2018}.  
Here as well, there have been several attempts to construct a continuum field theory to describe such non-equilibrium phase transition,  exploiting symmetry and conservation laws, or trying to make a smooth connection with the dynamics of the microscopic models \cite{stenhammar2013, Wittkowski2014,Nardini2017,Solon2018, Bialke2013, speck2014effective, farage2015effective, marconi2015towards, paoluzzi2020statistical} - in all cases under strong assumptions hard to put into test.

Besides the achiral active particles considered by the Vicsek and the ABP models, self-propelled chiral particles whose propulsion direction turns at a given rate, are also common to encounter at different scales, constituting yet another class of active particle systems. Examples   include microorganisms showing autonomous rotation, as is the case of \textit{E. Coli} near a wall \cite{berg1990,diluzio2005,lauga2006} and sperm cells \cite{Riedel2005,Friedrich2007}, L-shaped Janus colloids \cite{kummel2013circular}, or chiral grains \cite{arora2021emergent}, among others. 
The study of such  circle swimmers   from minimal models, usually thought of as extensions of the Vicsek and ABP models including an intrinsic frequency, is attracting increasing attention over the past few years \cite{lowen2016chirality, liebchen2017,levis2019,levis2019prr, Liao2018,Levis2018,bickmann2020chiral,lei2019,mani2022}, laying the groundwork for an analysis of the interplay between chirality and aligning interactions or excluded-volume effects, respectively. Systems of chiral active particles might feature both macrophase separation and microphase separation, depending on the rate at which their heading direction turns \cite{liebchen2017,Liao2018,mani2022}. 

The interplay between both excluded-volume and aligning interparticle interactions, or torques, has been addressed in a series of works \cite{farrell2012pattern,barre2015motility,Martin-Gomez2018,Sese-Sansa2018,Bhattacherjee2019,VanderLinden2019,Geyer2019,VanDamme2019,Liao2020,PeruaniRev,Jayaram2020,Zhang2021,Sese-Sansa2021,worlitzer2021motility}, focused on activity-induced  aggregation. 
Following the approach first introduced for pure ABP systems in \cite{Bialke2013}, a hydrodynamic description has been derived for specific types of polar and nematic (Vicsek-like) alignment rules \cite{Sese-Sansa2021}, for chiral ABP \cite{mani2022} and for dipolar ABP \cite{SeseSansaLiao}.
However, a general theoretical framework encompassing both  \textit{chiral} and  \textit{achiral} self-propelled particles  interacting via generic central forces and aligning torques, is still missing. 
Besides its formal interest, establishing a theory incorporating torques would allow to address several questions on general grounds. For instance, why, despite their fundamental difference, both circle swimmers and spinning particles (with no self-propulsion) suppress phase separation for large enough rotational frequencies. On the one hand, it is now known that intrinsic torques generically interrupt MIPS \cite{Liao2020,worlitzer2021motility,mani2022} in systems of chiral self-propelled particles, giving rise to clusters of self-limited size, both in systems with and without alignment interactions \cite{liebchen2017,levis2019,mani2022}. On the other hand, ferromagnetic colloidal particles, spinning at a given frequency imposed by an external magnetic field,  have a tendency to condense  as a result of mutual attractive interactions, but this phase separation is arrested at large enough spinning frequencies, giving rise, again, to finite-sized clusters \cite{massana2021prr}. To what extent a similar mechanism might explain finite-size clustering in both set-ups remains an interesting open problem. 

Here, we establish a general framework which allows to  systematically derive continuum hydrodynamic equations describing systems of self-propelled particles subjected to generic torques acting on the self-propulsion direction, being intrinsic to the particle or resulting from interparticle interactions. Interestingly, we can define a set of parameters that capture, at the microscopic level of pairwise interactions, the effect of both the activity and the torques considered. This paves the way for a mean-field analysis of the destabilisation of the homogeneous phase, which might lead to  different scenarios depending on the origin and type of torques considered. Within this framework, one can show that activity triggers a spinodal-like long-wavelength instability associated to MIPS (the location of which is affected by the self-propulsion mechanism and the different interactions), while self-rotations trigger a short-wavelength instability, introducing a characteristic length $\ell$. This characteristic length appears to scale as the inverse of the turning rate  $\ell\sim\omega^{-1}_0$,  as found in a simple model of chiral active particles with polar alignment \cite{liebchen2017} and in suspensions of spinning hematite colloids \cite{massana2021prr}, suggesting a common mechanism underlying these two \textit{a priori} unrelated physical systems.

The paper is organised as follows. In \cref{deriv_hydro_eq}, starting from the microscopic dynamics, we derive the (mean-field) continuum hydrodynamic equations that govern the evolution of the density and the polarization fields. We then go on to analyse the linear stability of the homogeneous and isotropic phase in \cref{stabanaly}. We formally show that adding both an intrinsic frequency of rotation and interparticle torques to the dynamical equations qualitatively changes the phase behavior predicted by the linear stability analysis. We discuss the different cases in \cref{longwavelengthinst,shortwavelengthinst}.




\section{Derivation of the continuum equations}\label{deriv_hydro_eq}

We consider a system of self-propelled particles governed by the following $N$-body Smoluchowski equation, accounting for the time evolution of the joint probability distribution function $\psi_{N}(\Gamma=\{\textbf{r}_{i}, \varphi_{i}\}_{i=1..N},t)$,

\begin{equation}
\begin{split}
\partial_{t} \psi_{N}& = \sum_{i=1}^{N} \nabla_{i} \cdot \left[ (\nabla_{i} \Ua) \psi _{N} - v_{0} \textbf{e}_{i} \psi _{N} + D_0\nabla_{i} \psi _{N} \right]  +\\
 & \qquad  \sum_{i=1}^{N} \partial _{\varphi_{i}} \left[ \left(\partial_{\varphi_{i}} \Ua \right) \psi _{N}- \omega_0\psi _{N} +D_{r} \partial_{\varphi_{i}} \psi_{N} \right].
\label{derivat-hydro_eq:0}
\end{split}
\end{equation}
The function $\psi_{N}$ gives the probability to find $N$ particles of the system at $N$ given positions in a 2D space, $\textbf{r}_{i}(t) = (x_{i},y_{i})$, and with $N$ given orientations $\varphi_i$.
The system is composed of particles that self-propel at constant speed, $v_0$, along $ \textbf{e}_{i} = (\cos \varphi_i, \sin \varphi_i )$,  {and rotate at an intrinsic frequency, $\omega_0$}. They are also subjected to thermal and rotational noise, characterised by the diffusion constants $D_0$ and $D_r$, respectively. From now on, we set $D_0=1$ without loss of generality.

Interactions are modelled by the pairwise interaction potential
\begin{equation}
\begin{split}
 \Ua\left(\left\{ \textbf{r}_j \right\}, \left\{ \varphi_j \right\} \right) = \sum_{i=1}^{N} \sum_{i < j} u (|\textbf{r}_{j} - \textbf{r}_i|,  \varphi_{i}, \varphi_j).
 \label{derivat-hydro_eq:00}
\end{split}
\end{equation}
For the sake of generality, we do not specify the functional form of $\Ua$. Consequently, we are deriving a  framework to describe systems of self-propelled particles whose interactions depend on the center-to-center distance between pairs and on their inner orientation.

Integrating out the degrees of freedom of $(N-1)$ particles, one can obtain the 1-body Smoluchowski equation
\begin{equation}
\begin{split}
\partial_{t} \psi_{1} &= - \nabla_{1} \cdot \left[ \textbf{F} \left(\textbf{r}_{1},\varphi_{1};t\right)+ v_{0} \textbf{e}_1 \psi _{1} - \nabla_{1} \psi _{1} \right] \\
&\qquad  - \frac{\partial}{\partial \varphi_{1}} \left[ T \left(\textbf{r}_{1},\varphi_{1};t \right) + \omega_0\psi _{1} - D_{r} \frac{\partial \psi_{1}}{\partial \varphi_{1}} \right],
 \end{split}
\label{derivat-hydro_eq:2}
\end{equation}
which constitutes the first equation of a BBGKY-like hierarchy of equations, coupled to two-body terms through $\textbf{F} \left(\textbf{r}_{1},\varphi_{1},t\right)$ and $T \left(\textbf{r}_{1},\varphi_{1},t \right)$, which are the effective force and torque, respectively, exerted by the surrounding particles into the tagged particle (labeled \textit{1}). The effective force reads,

\begin{equation}
\begin{split}
\textbf{F} (\textbf{r}_{1},\varphi_{1},t)&= -N \int_{-\infty}^{\infty} d\textbf{r}_{2} ... d\textbf{r}_{N} \int_{0}^{2\pi} d\varphi_{2} ... d\varphi_{N} \nabla_{1} \Ua \: \psi _{N}\\
&= -\int_{-\infty}^{\infty} d\textbf{r}_{2} \int_{0}^{2\pi} d\varphi_{2} \nabla_1 u (|\textbf{r}_{2} - \textbf{r}_1|,  \varphi_{1}, \varphi_2)  \psi _{2}
\end{split}
\label{appx_continuum_model_forcetorque_eq:1}
\end{equation}
and the effective torque is, 
\begin{equation}
\begin{split}
T(\textbf{r}_{1},\varphi_{1},t) &=-N \int_{-\infty}^{\infty} d\textbf{r}_{2} ... d\textbf{r}_{N} \int_{0}^{2\pi} d\varphi_{2} ... d\varphi_{N} \left(\partial_{\varphi_{1}}\Ua \right)\psi _{N} \\
&=- \int_{-\infty}^{\infty} d\textbf{r}_{2}  \int_{0}^{2\pi} d\varphi_{2}   \partial_{\varphi_{1}} u (|\textbf{r}_{2} - \textbf{r}_1|,  \varphi_{1}, \varphi_2) \psi _{2}
\end{split}
\label{appx_continuum_model_forcetorque_eq:2}
\end{equation}
where  $\psi_{2}(\textbf{r}_{1}, \textbf{r}_{2}, \varphi_{1}, \varphi_{2},t)$  is the two-body probability distribution.

Forces come from the spatial dependency of the pair potential. Typically, one considers excluded-volume interactions, which set the particles' characteristic finite size. Conversely, torques result from the orientation dependency in \cref{derivat-hydro_eq:00} and thus act on the direction of self-propulsion of particles. Depending on the type of aligning potential considered, torques can lead to different scenarios. Some particular cases have been already studied in the literature.
Vicsek-like aligning rules, which are decoupled from the excluded-volume forces, are considered in \cite{Sese-Sansa2021}. On the contrary, dipolar interactions between permanent point dipoles, which couple spatial and angular degrees of freedom, are analysed in \cite{SeseSansaLiao}.  

To proceed, we introduce a change of variables. The system we are considering is composed of a pair of particles and can be fully defined by the vector distance  $\textbf{r}_{12} = \textbf{r}_{2} - \textbf{r}_1 = r_{12}(\cos \omega, \sin \omega)$ and the orientations $\varphi_1$, $\varphi_2$. Thus, in the lab frame of reference, the set of variables is  $(r_{12},\omega,  \varphi_{1}, \varphi_2) $. We note that the integrals in \cref{appx_continuum_model_forcetorque_eq:1,appx_continuum_model_forcetorque_eq:2} are over $\textbf{r}_{2} $ and $\varphi_2$, while  $\textbf{r}_{1} $ and $\varphi_1$ are kept fixed. This allows us to express orientations as a function of $\varphi_1$, and therefore we define $\varphi_{12} = \varphi_2 - \varphi_1$. Employing a body-fixed frame, one can express the directions along the plane relative to $\textbf{e}_{1}$. We thus introduce $\theta_1=\varphi_1-\omega$ and $\theta_2=\varphi_2-\omega$. However, $\theta_2$ can be expressed as a function of $\theta_1$ and $\varphi_{12}$, $\theta_2=\varphi_{12}-\theta_1$. We can therefore use $r_{12}$, $\theta_1$ and $\varphi_{12}$ as our set of independent variables, without loss of generality.

We now decompose $\psi_{2}$, in terms of the new set of variables, using the following identity 
\begin{equation}
\psi_{2}(\textbf{r}_1,\textbf{r}_2,\varphi_{1},\varphi_{2},t) =\bar{\rho}\, \psi_{1}(\textbf{r}_1,\varphi_{1},t)\,  \mathcal{G}(r_{12},\theta_1,\varphi_{12},t)
\label{theoret-mean-field_eq:5}
\end{equation}
where $\bar{\rho}$ is the average density and $\mathcal{G}(r_{12},\theta_1,\varphi_{12},t)$ the pair correlation function encoding the microscopic structure of the system. We interpret it as the probability of finding a particle with orientation $\varphi_{2}$ in the plane-direction $\theta_1$, at a distance $ r_{12} = |\textbf{r}_{12} |$ from the tagged particle (at  $\textbf{r}_{1}$  with orientation $\varphi_{1}$).

We also introduce the change of variables $ d  \textbf{r}_{12} =d\textbf{r}_{2} $ and $ d  \varphi_{12} =d\varphi_{2} $, stemming from the definition $\textbf{r}_{12} = \textbf{r}_{2} - \textbf{r}_{1}$ and  $\varphi_{12} = \varphi_{2} - \varphi_{1}$. This yields the rewriting of \cref{appx_continuum_model_forcetorque_eq:1,appx_continuum_model_forcetorque_eq:2} as,

\begin{equation}
\begin{split}
\textbf{F}\left(\textbf{r}_{1},\varphi_{1};t\right)&=\bar{\rho} \psi_{1}\left(\textbf{r}_1,\varphi_{1},t\right)\int_{-\infty}^{\infty} d\textbf{r}_{12} \\
&\qquad \int_{0}^{2\pi}d\varphi_{12}
\frac{\partial u \left(\textbf{r}_{12}, \varphi_{12} \right)}{\partial r_{12} } \frac{\textbf{r}_{12}}{r_{12}}  \mathcal{G}\left(r_{12},\theta_1,\varphi_{12},t\right),
\label{appx_continuum_model_forcetorque_eq:3}
\end{split}
\end{equation}
\begin{equation}
\begin{split}
T \left(\textbf{r}_{1},\varphi_{1};t\right) &=  \bar{\rho} \psi_{1}\left(\textbf{r}_1,\varphi_{1},t\right) \int_{-\infty}^{\infty} d\textbf{r}_{12}\\
&\qquad \int_{0}^{2\pi} d\varphi_{12} \partial_{ \varphi_{12}} u \left(\textbf{r}_{12}, \varphi_{12} \right)  \mathcal{G}\left(r_{12},\theta_1,\varphi_{12},t\right)
\label{appx_continuum_model_forcetorque_eq:4}
\end{split}
\end{equation}
In the remainder of the paper, we shall drop the subscripts for clarity. We will now group the two-body terms in the force and torque's expression in single scalar coefficients.

\paragraph{Torque} Grouping the two-body terms in the torque's expression in a scalar coefficient, $\kappa$, allows for the rewriting of \cref{appx_continuum_model_forcetorque_eq:4} as,

\begin{equation}
\begin{split}
T = -\bar{\rho} \psi_{1} \kappa,
\end{split}
\label{modeldescrip_eq:13}
\end{equation}
where
\begin{equation}
\begin{split}
\kappa= -\int_{0}^{\infty} dr  r  \int_{0}^{2\pi} d\theta  \int_{0}^{2\pi} d\varphi \partial_{ \varphi} u \left(\textbf{r}, \varphi \right)  \mathcal{G}\left(r,\theta,\varphi,t\right) \,.
\end{split}
\label{modeldescrip_eq:12}
\end{equation}
This factor $\kappa$ is linked to the spatial and orientational correlations encoded in $\mathcal{G}\left(r,\theta,\varphi,t\right)$. A homogeneous suspension of passive particles $(v_0=0)$ is spatially and orientationally uncorrelated. It is equally probable to find a particle at any distance from the tagged particle, $r$, in any in-plane direction, $\theta$ and with any relative orientation $\varphi$. Therefore, the correlation function fulfils the \textit{head-tail} symmetry ($\theta \rightarrow \theta + \pi$) and the symmetry against \textit{exchange of particles' position} ($\theta \rightarrow - \theta$), as well as the \textit{parallel-antiparallel} symmetry ($\varphi \rightarrow \varphi +\pi$) and the symmetry against \textit{exchange of orientations} ($\varphi \rightarrow - \varphi$). Introducing activity breaks the symmetry $\theta \rightarrow \theta + \pi$. This implies that it is more likely to find another particle in front of the tagged particle than behind of it, a signature of the self-trapping mechanism that leads to MIPS. 

Keeping these symmetries in mind, it is straightforward to argue the cases in which $\kappa$ has a non-zero value. The alignment mechanisms usually studied in the field enter in the interaction potential with an even dependency in $\varphi$ (e.g. Vicsek-like alignment interactions). Alternatively, one may think of more complex interactions also leading to effective alignment, like dipole-dipole interactions, which also involve an even dependency in $\theta$ (i. e. dipolar interactions involve a dependency on both relative orientations and relative positions in space). All in all, this results in $\partial_{ \varphi} u \left(\textbf{r}, \varphi \right)$ having an odd dependency in $\varphi$ and, depending on the specific interaction considered, also in $\theta$. Thus, the product of $\partial_{ \varphi} u \left(\textbf{r}, \varphi \right)$ times a correlation function $\mathcal{G}\left(r,\theta,\varphi,t\right)$, which fulfils the symmetries against \textit{exchange of particles' position} and \textit{exchange of orientations} (i. e., it is even in $\varphi$ and $\theta$), results in $\kappa =0$ upon integration, \cref{modeldescrip_eq:12}. This scenario does not change in the presence of activity. We thus state that $\kappa$ remains identically zero as long as the symmetry $\varphi \rightarrow - \varphi$ and/or $\theta \rightarrow - \theta$ are not broken. 

\paragraph{Force} In the case of \cref{appx_continuum_model_forcetorque_eq:3}, it is not straightforward to group the two-body terms in scalar coefficients. To do so, we first decompose $\textbf{F} (\textbf{r}_{1},\varphi_{1};t)$ in the vector basis spanned by the direction of self-propulsion and the gradient of the probability density, $(\textbf{e}, \nabla \psi_{1})$. We follow a Gram-Schmidt orthonormalization scheme (see \cref{appx_gram_schmidt} for the full derivation), which in this case is an approximation, due to the fact that it is not guaranteed that $\textbf{e}$ and $\nabla \psi_{1}$ remain linearly independent, since they evolve in time and could become, at some point, parallel throughout the system's evolution. The decomposition of $\textbf{F} (\textbf{r}_{1},\varphi_{1};t)$ reads,

\begin{equation}
\textbf{F} \approx \left(-\bar{\rho} \psi_{1} \zeta \right )\textbf{e} + \left(1-\Da \right)\nabla\psi_{1},
\label{theoret-mean-field_eq:10}
\end{equation}
where the two scalar coefficients introduced correspond to,

\begin{equation}
\begin{split}
&\zeta=-\int_{0}^{\infty} dr \: r \int_{0}^{2\pi} d\theta \cos \theta\int_{0}^{2\pi} d\varphi  \frac{\partial u \left(\textbf{r}, \varphi \right)}{\partial r }  \mathcal{G}\left(r,\theta,\varphi,t\right)
\end{split}
\label{theoret-mean-field_eq:7}
\end{equation}
and 
\begin{equation}
\Da=1-  \frac{ \left( \nabla \psi_1 - \left( \textbf{e} \cdot \nabla \psi_1 \right) \textbf{e} \right) \cdot   \textbf{F} }{|\nabla \psi_1|^2}.
\label{modeldescrip_eq:14bis}
\end{equation} 
The first term on the right-hand side (RHS) of \cref{theoret-mean-field_eq:10} is the component of the force acting along the direction of self-propulsion. We can interpret this component as the one quantifying the imbalance between the self-propulsion of the tagged particle and its arrest induced by collisions with neighbouring particles. 

In a system of passive colloids $\zeta =0$, which can be simply understood applying the same symmetry arguments given earlier and thus, it is independent of the interparticle potential. This further means that, in the present construction, $\zeta$ is irrelevant for a standard spinodal decomposition in an equilibrium system of attractive particles at low enough temperatures. Our approach is particularly tailored for activity-induced aggregation, exploiting the preferred direction of motion $\textbf{e}$  to decompose the effective force \cref{appx_continuum_model_forcetorque_eq:3}. In order to account for equilibrium phase separation, one could invoque a mean-field approximation and split the two-body distribution function as a product of one-body ones. This will lead to an effective diffusivity at the level of the one-body Smoluchowski equation (see below) that will change sign when the homogenous state becomes unstable, signalling a spinodal long-wavelength instability. 


As soon as activity enters the systems, the $\theta \rightarrow \theta + \pi $ symmetry is broken and thus $\zeta \neq 0$, due to the $\cos \theta$ term stemming from the projection of the force (see the Gram-Schmidt orthonormalization in \cref{appx_gram_schmidt}). We note that in the active case $\zeta$ will also have contributions from the aligning potential, evidencing that in the model we have derived, alignment interactions modify the force imbalance arising from the collision persistence and captured by $\zeta$.

The second term on the RHS of \cref{theoret-mean-field_eq:10} can be interpreted as an effective diffusion acting along the gradient of the one-body probability distribution.

Introducing the expressions for the force and the torque, \cref{modeldescrip_eq:13,theoret-mean-field_eq:10}, into the one-body Smoluchowski equation leads to the rewriting of \cref{derivat-hydro_eq:2} as, 

\begin{equation}
\begin{split}
\partial_{t} \psi_{1} = - \nabla \cdot \big[ v_{\bar{\rho}} \,\textbf{e} \psi_{1}- \Da \nabla\psi_{1} \big] - \frac{\partial}{\partial \varphi} \big[ \varepsilon_{\bar{\rho}} \psi _{1} - D_{r} \frac{\partial\psi_{1}}{\partial \varphi} \big],
\end{split}
\label{modeldescrip_eq:14}
\end{equation}

where 

\begin{align}
v_{\bar{\rho}} = v_0 -\bar{\rho} \zeta,              &&
\varepsilon_{\bar{\rho}}= \omega_0-\bar{\rho} \kappa.
\label{modeldescrip_eq:14.1}
\end{align}

The first two terms on the RHS of \cref{modeldescrip_eq:14} correspond to the advection and diffusion, respectively, of the spatial degrees of freedom. Here, the translational advection term sets an effective self-propulsion speed, $v_{\bar{\rho}}$, \cref{modeldescrip_eq:14.1}, which decays with the mean density, $\bar{\rho}$, at a rate given by $\zeta$ and which can thus be interpreted as a \textit{translational friction coefficient}, accounting for the arrest of particles in crowded environments. $\Da$ is an effective many-body diffusivity. 
The third and fourth terms on the RHS of \cref{modeldescrip_eq:14} correspond to the advection and diffusion of the orientations. In the advective term, $ \varepsilon_{\bar{\rho}}$ is the effective frequency of rotation, where $\omega_0$ is the intrinsic frequency of rotation, as stated earlier, and $\kappa$ can be interpreted, by analogy with $\zeta$, as a \textit{rotational friction coefficient}, stemming from interparticle aligning torques.  Note the equivalent role played by ($v_{\bar{\rho}}$, $v_0$, $\zeta$) and  ($ \varepsilon_{\bar{\rho}}$, $\omega_0$, $\kappa$),  \cref{modeldescrip_eq:14.1}.  Finally, the rotational diffusion coefficients is $D_{r}$.


The microscopic information of the one-body equation just derived is captured by $\zeta$, $\kappa$ and $\Da$, which link the one-body distribution to higher order ones.
To proceed, we make the assumption that $\zeta$, $\kappa$ and $\Da$ are independent of the tagged particle's position, which is valid as long as the system is in (close to)  an homogeneous state. 
We therefore close the hierarchy of coupled equations by considering these parameters as constants.
This is a central approximation of our approach, which allows us to derive  effective hydrodynamic equations. 

We emphasize that, opposed to top-down approaches that base the derivation of the effective hydrodynamic equations on symmetry arguments and conservation laws \cite{Wittkowski2014}, our approach directly coarse-grains the microscopic dynamics. Thus, the coefficients we define are not phenomenological but stem from interparticle interactions. They indeed take specific numerical values in particle-based models, whose calculation allows for a direct quantitative comparison between the microscopic model and the coarse-grained theory. Another relevant feature of our approach is the effective speed $v_{\bar{\rho}}$ decaying at increasing density (a signature of MIPS), which here is an outcome of the derivation and not introduced as an hypothesis \cite{Bialke2013}.

We can now derive the hydrodynamic equations by integrating the closed one-body Smoluchowski equation, \cref{modeldescrip_eq:14}. We define the first two moments of the one-body probability distribution to be the density field 
\begin{equation}
\rho (\textbf{r},t) \equiv \int_{0}^{2 \pi} d \varphi \psi_{1} (\textbf{r},\varphi,t),
\end{equation}
and the polarization 
 \begin{equation}
 \textbf{p} (\textbf{r},t) \equiv \int_{0}^{2 \pi} d \varphi \textbf{e} \psi_{1} (\textbf{r},\varphi,t),
 \end{equation}
 which lead to the hydrodynamic equations
\begin{equation}
\begin{split}
 \partial_{t} \rho (\textbf{r},t) =- \nabla \cdot \Big[ v_{\bar{\rho}} \textbf{p} -\Da \nabla \rho \Big],
\end{split}
\label{modeldescrip_eq:15}
\end{equation}
\begin{equation}
\begin{split}
 \partial_{t} \textbf{p} (\textbf{r},t) = -\nabla \cdot \Big[v_{\bar{\rho}} (\frac{1}{2} \rho \mathbb{1} + \textbf{Q} )- \Da \nabla \textbf{p} \Big] 
 -  \varepsilon_{\bar{\rho}} \textbf{p}^{\perp} - D_{r} \textbf{p}. 
 \end{split}
\label{modeldescrip_eq:16}
\end{equation}
Here, $\perp$ indicates a rotation corresponding to $\bold{p}^{\perp}=\mathcal{R}\bold{p}$, $\nabla^{\perp}=\mathcal{R}\nabla$ with  $\mathcal{R} = \left(\begin{array}{ccc} 0 & -1\\1 & 0\\ \end{array}\right)$.  The time evolution equation of each moment is linearly coupled to the next order moment. Therefore, the time evolution of the polarization is coupled to the nematic tensor, $\bold{Q}$. To close the set of hydrodynamic equations, we drop the dependency of \cref{modeldescrip_eq:16} on $\bold{Q}$. As we show in \cref{alternative_adiabatic}, after performing an adiabatic approximation to the hydrodynamic equations (i. e. $ \partial_{t} \textbf{p}=0$) we still capture the relevant information on the destabilization modes at any wave vector. This proves that $\rho (\textbf{r},t)$ is the slowest moment of the probability distribution and the higher order moments are enslaved to it. This, in turn, justifies cutting the hierarchy of hydrodynamic equations to $\bold{Q}$.

The hydrodynamic equations above have an isotropic homogenous steady-solution $(\rho (\textbf{r},t)=\bar{\rho}, \textbf{p} (\textbf{r},t) =\textbf{0})$ but do not admit a polar steady-solution, as continuum theories of flocking. The present theory does not provide a symmetry breaking term \textit{\`a la } Landau, as in the Toner-Tu theory, and therefore, it is limited to the description of non-polar states. 
\section{Linear stability analysis}\label{stabanaly}
We now assume that the density $\rho(\textbf{r},t)$ is a slowly varying field \cite{Bialke2013}. This approximation is needed in order to observe a linear instability and it is justified as long as we are interested in the stability of a homogeneous isotropic state. In this case, one can replace $\bar{\rho}$ by the local density field $\rho(\textbf{r},t)$ in the hydrodynamic equations, \cref{modeldescrip_eq:15,modeldescrip_eq:16}. 
Thus, the closed set of hydrodynamic equations accounting for the time evolution of a perturbation around the homogeneous and isotropic state, $\rho (\textbf{r},t) =\bar{\rho} +\delta \rho$ and $\textbf{p} (\textbf{r},t)=\delta \textbf{p}$ is

\begin{equation}
\begin{split}
\partial_{t} \delta \rho  =- \nabla \cdot \Big[ (v_0 - \bar{\rho} \zeta) \delta\textbf{p} -\Da \nabla \delta\rho \Big],
\end{split}
\label{linear-stab-analys_eq:1}
\end{equation}

\begin{equation}
\begin{split}
\partial_{t} \delta \textbf{p} = -\nabla \cdot \Big[\frac{1}{2}  \big(v_0 -2 \bar{\rho} \zeta \big) \delta \rho  - \Da \nabla \delta \textbf{p} \Big]- D_r   \varepsilon_{\bar{\rho}}  \delta\textbf{p}^{\perp} - D_{r} \delta \textbf{p}.
\end{split}
\label{linear-stab-analys_eq:2}
\end{equation}

Our goal is to study how torques affect the structure formation in systems of self-propelled particles, which result from the competition between self-propulsion and interparticle collisions. In the mean-field model we have derived, the effect of torques (intrinsic or due to interparticle alignment) is captured in $\varepsilon_{\bar{\rho}}$, while $\zeta$ quantifies the collision persistence. Thus, to explore the phenomenology of our model we need to scan a set of three parameters: $v_0$, $\varepsilon_{\bar{\rho}}$ and  $\zeta$. From now on, we  note $\varepsilon_{\bar{\rho}}=\varepsilon$.


It is possible to write the hydrodynamic equations for the perturbation in Fourier space, $ \textbf{u} \sim \hat{\textbf{u} }e^{i \textbf{q} \cdot \textbf{r}}$, where $\textbf{u}=\left(\delta \rho, \delta \textbf{p} \right)$, which finally leads to

\begin{equation}
\begin{split}
\partial_{\tilde{t}} \delta \hat{ \tilde{\rho}}  =- i \tilde{\textbf{q}} \cdot \Big[ 4(\frac{v_0}{ v^*} - \tilde{\zeta}_{0}) \delta\hat{\tilde{\textbf{p}}} -i \tilde{\textbf{q}}\delta\hat{\tilde{\rho}} \Big],
\end{split}
\label{linear-stab-analys_eq:5}
\end{equation}

 \begin{equation}
\begin{split}
\partial_{\tilde{t}} \delta \hat{\tilde{\textbf{p}}} = -  i \tilde{\textbf{q}} \cdot \big[2 (\frac{v_0}{ v^*} -2 \tilde{\zeta}) \delta \hat{\tilde{\rho}} -  i \tilde{\textbf{q}} \delta \hat{\tilde{\textbf{p}}} \big]  -  \tilde{\varepsilon}  \delta\hat{\tilde{\textbf{p}}} ^{\perp}  -\delta \hat{\tilde{\textbf{p} }},
\end{split}
\label{linear-stab-analys_eq:6}
\end{equation}

where the dimensionless quantities read,

\begin{align}
\tilde{t}=D_{r}t,               &&
\tilde{\textbf{q}}=\sqrt{\frac{\Da}{D_{r} }}\textbf{q},\\ \nonumber
\frac{v_0}{v^*}=\frac{v_0}{4\sqrt{\Da D_{r} } },      &&
\tilde{\varepsilon} = \frac{\varepsilon}{D_r}, &&
\tilde{\zeta} = \frac{\bar{\rho}}{v^*}\zeta.  
\label{linear-stab-analys_eq:4}
\end{align}
In the remainder of the paper, we work with dimensionless quantities but we drop the tilde $\tilde{\textbf{q}}\equiv{\textbf{q}}$. 
Writing the two dimensionless independent linearized equations in matrix form, $\partial_{t}(\delta \hat{\rho} \;   \delta \hat{\textbf{p}}   )^{T} = M (\delta \hat{\rho} \;   \delta \hat{\textbf{p}} )^{T}$, where

 \[
  M =
   \begin{bmatrix}
  -\textbf{q}^2 & - i 4(\frac{v_0}{ v^*} - \zeta) q_x & - i 4(\frac{v_0}{ v^*} - \zeta) q_y  \\
    - i 2 (\frac{v_0}{ v^*} -2 \zeta) q_x & - (\textbf{q}^2+1 ) & \varepsilon\\ 
    - i 2 (\frac{v_0}{ v^*} -2 \zeta) q_y & -\varepsilon& - (\textbf{q}^2+1 )
  \end{bmatrix},
\]
we can compute the system's eigenvalues by solving the determinant of $M$ and setting it to 0. The details of the computation as well as the functional form of the eigenvalues can be found in \cref{eigenval_compu}. 

The eigenvalues correspond to the dispersion relations quantifying the growth of a perturbation with dimensionless wave vector $\textbf{q}$, and allow us to explore the  onset of linear instabilities. As mentioned before, the parameter space of our model is conformed by $\frac{v_0}{v*}$, $\zeta$, $\varepsilon$. In \cref{sectionepsilon0}, we briefly discuss the torque-free case $\varepsilon =0$, which has been extensively studied in \cite{Bialke2013,Speck2015} and which we add for completeness. Here, the two relevant parameters controlling the system's destabilization are $\frac{v_0}{v*}$ and $\zeta$. We then move on to introduce $\varepsilon$ and show that the predictions of the linear stability analysis qualitatively change in the presence of \textit{effective torques}.

To obtain $\varepsilon \neq 0$ one can think of a functional form of the alignment potential involving odd dependencies in the angular variables, which would automatically lead to a $\kappa \neq 0$ upon integration. It is worth mentioning, though, that alignment interactions leading to either parallel or antiparallel alignment involve even dependencies in the angular variables. Alternatively, one can consider a nonreciprocal pairwise alignment interaction, that breaks the symmetry under \textit{exchange of orientations} $\varphi \rightarrow - \varphi$ and/or under \textit{exchange of particles' position} $\theta \rightarrow - \theta$, and which thus results in $\kappa \neq 0$. Besides, chiral active particles self-rotate at an intrinsic frequency $\omega_0$, adding a constant (non-zero) contribution to $\varepsilon$. 

 

\subsection{Case $\varepsilon = 0$ }\label{sectionepsilon0}

For systems with no effective torques, $\varepsilon =0$. In this case,  originally considered in \cite{Bialke2013}, the phase behavior is exclusively controlled by the competition between activity and interparticle collisions. Alternatively, effective torques can also lead to $\varepsilon =0$ in some particular microscopic achiral models, such as systems with Vicsek-like alignment  \cite{Sese-Sansa2021} or  dipole-dipole interactions \cite{SeseSansaLiao}, where the symmetry of the alignment interaction together with the symmetry of the angular correlation leads to a rotational friction coefficient that is identically zero upon integration (see discussion on angular symmetries in \cref{deriv_hydro_eq}). 

Setting $\varepsilon = 0$, the eigenvalues (see \cref{linear-stab-analys_eq:9} in \cref{eigenval_compu}) can be written as
\begin{equation}
\begin{split}
\lambda_{1}&=-\frac{1 }{2}(2 \textbf{q}^2 + 1)  + \sqrt{1 - 32\textbf{q}^2  \Big(\frac{v_0}{v^*}-\zeta\Big)\Big(\frac{v_0}{v^*}-2\zeta\Big)},\\
\lambda_{2}&=-( {\textbf{q}}^2 +1),\\
\lambda_{3}&=-\frac{1 }{2}(2 \textbf{q}^2 + 1)  - \sqrt{1 - 32\textbf{q}^2  \Big(\frac{v_0}{v^*}-\zeta\Big)\Big(\frac{v_0}{v^*}-2\zeta\Big)}.
\end{split}
\label{computeigenval_eq:4}
\end{equation}
\begin{figure}[h]
	 \includegraphics[trim=180 70 0 265, width=\columnwidth]{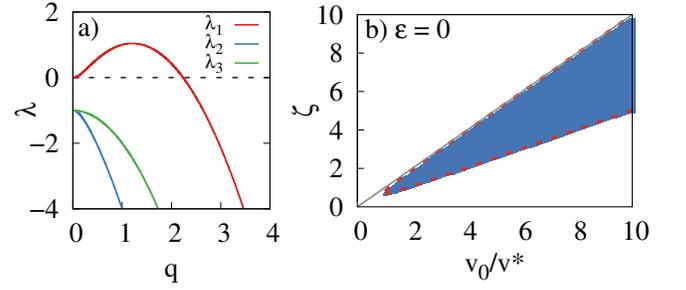}
	\caption{ a) Dispersion relations as a function of the dimensionless wave vector $q =|\textbf{q}|$ for $v_0/v^* = 2.5$, $\zeta = 2.0$ and  $\varepsilon = 0.0$: $\lambda_2<0$ and $\lambda_3<0$ for all $q$, while $\lambda_1(q)>0$ at small wavenumbers. b) LW instability region in the $(\frac{v_0}{ v^*}, \zeta)$ plane for $\varepsilon = 0$. The limit of stability illustrated by the red dashed line corresponds to \cref{computeigenval_eq:6}.}
	\label{fig1_reginstab_q0_eps0}
	\end{figure}
We now focus our attention on the low-$q$ behavior of these eigenvalues which, to 2nd order in $q =|\textbf{q}|$,  read
\begin{equation}
\begin{split}
\lambda_{1}&= 0  -\left[1 + 8  \left(\frac{v_0}{v^*}-\zeta\right)\left(\frac{v_0}{v^*}-2\zeta\right)\right]\textbf{q}^2 + \mathcal{O}(\textbf{q}^3),\\
\lambda_{2}&=-1- \textbf{q}^2, \\
\lambda_{3}&=-1  +\Big[-1 + 8  \Big(\frac{v_0}{v^*}-\zeta\Big)\Big(\frac{v_0}{v^*}-2\zeta\Big)\Big]\textbf{q}^2 + \mathcal{O}(\textbf{q}^3).
\end{split}
\label{computeigenval_eq:5}
\end{equation}
The eigenvalues $\lambda_{2}$ and $\lambda_{3}$ are  negative for any wave vector $\textbf{q}$, indicating that the homogeneous state is stable upon a perturbation along these two modes, as depicted in \cref{fig1_reginstab_q0_eps0} a). On the contrary, $\lambda_{1}$ can become positive at $\textbf{q} \rightarrow 0$, and trigger the growth of a long-wavelength (LW) instability. The instability region at $\textbf{q} \rightarrow 0$ as a function of $(\frac{v_0}{v^*},\zeta )$ is represented in \cref{fig1_reginstab_q0_eps0} (b) in blue. One can obtain the limits of stability analytically by taking the second order term in the Taylor expansion, \cref{computeigenval_eq:5}, and setting it to zero, leading to
\begin{equation}
\zeta(\frac{v_0}{v^*}, \varepsilon=0,\textbf{q} \rightarrow 0)=\zeta_0 = \frac{3}{4}\frac{v_0}{v^*} \pm \frac{1}{4}\sqrt{\Big(\frac{v_0}{v^*}\Big)^2-1},
\label{computeigenval_eq:6}
\end{equation}
and represented by two broken lines in \cref{fig1_reginstab_q0_eps0} (b). 
We note that the dispersion relations do not have any complex term, implying that no oscillating instabilities take place.

Such LW instability, coming from an increase of the effective friction $\zeta$ along the direction of self-propulsion as the self-propulsion speed increases, is associated to MIPS \cite{Bialke2013}. 

\subsection{Case $\varepsilon \neq 0$}\label{sectionepsilonnot0}

We now study the impact of a non-vanishing $\varepsilon$ in the stability of the homogeneous phase. We thus explore the parameter space $(\frac{v_0}{ v^*}, \zeta, \varepsilon)$ and show that taking $\varepsilon$ into account qualitatively changes the linear stability of the homogeneous isotropic state. We recall that one way of realising this is by considering self-turning chiral particles, as detailed in \cref{deriv_hydro_eq}.

 
\subsubsection{Long-wavelength instabilities}\label{longwavelengthinst}
\begin{figure}[h]
	\centering
	 \includegraphics[trim=180 70 0 265, width=\columnwidth]{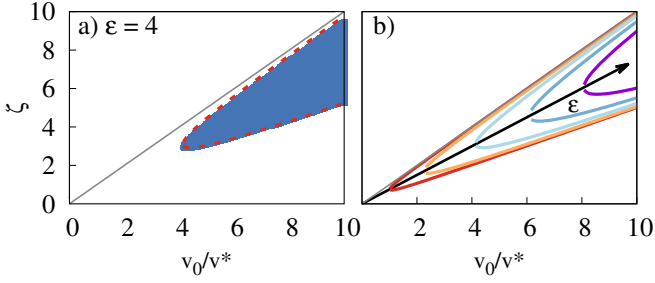}
	\caption{a) Long-wavelength instability region  for $\varepsilon = 4$. The limit of stability illustrated by the red dashed line corresponds to \cref{caseepsil_eq:1}. 
	b) Long-wavelength  limit of stability given by \cref{caseepsil_eq:1} for $\varepsilon =1, 2, 4, 6, 8$, and showing the shift of the unstable region to higher values of $\frac{v_0}{v^*}$ as $\varepsilon$ increases. }
	\label{fig1_reginstab_q0}
	\end{figure}
	
We start our study analysing LW instabilities ($\textbf{q}\rightarrow 0$), which signal the formation of a macroscopic structure. In the model we present, \cref{modeldescrip_eq:15,modeldescrip_eq:16}, such LW instability is associated to a phase separation identified with MIPS. 
The expansion of the eigenvalues, \cref{linear-stab-analys_eq:9} in \cref{eigenval_compu}, up to 2nd order in $\textbf{q}$ leads to,

\begin{equation}
\begin{split}
\lambda_{1}&= 0  -\left[1 + \frac{8}{1+\varepsilon^2} \left(\frac{v_0}{v^*}-\zeta\right)\left(\frac{v_0}{v^*}-2\zeta\right)\right]\textbf{q}^2 + \mathcal{O}(\textbf{q}^3)\\
\lambda_{2}&=-1 - i \varepsilon + \left[-1 +  \frac{4}{1+\varepsilon^2}\left(\frac{v_0}{v^*}-\zeta\right)\left(\frac{v_0}{v^*}-2\zeta\right)\right. \\
&\qquad \left.- i \frac{4\varepsilon}{1+\varepsilon^2}\left(\frac{v_0}{v^*}-\zeta\right)\left(\frac{v_0}{v^*}-2\zeta\right) \right] \textbf{q}^2 +  \mathcal{O}(\textbf{q}^3)\\
\lambda_{3}&=-1 + i \varepsilon + \left[-1 +  \frac{4}{1+\varepsilon^2}\left(\frac{v_0}{v^*}-\zeta\right)\left(\frac{v_0}{v^*}-2\zeta\right) \right. \\
&\qquad \left. + i \frac{4\varepsilon}{1+\varepsilon^2}\left(\frac{v_0}{v^*}-\zeta\right)\left(\frac{v_0}{v^*}-2\zeta\right) \right] \textbf{q}^2 +  \mathcal{O}(\textbf{q}^3).\\
\end{split}
\label{caseepsil_all_eq:1}
\end{equation}
Out of the three eigenvalues, only $\lambda_1$ gives rise to a growing instability, just as in the $\varepsilon =0$ case, \cref{sectionepsilon0}. Moreover, we note that  $\lambda_1$  does not have a complex part.

We investigate the LW instability region in the $(\frac{v_0}{v^*},\zeta)$ plane by numerically solving  the 'full' dispersion relation $\lambda_1$ (given in \cref{eigenval_compu}, \cref{linear-stab-analys_eq:9}) for different $\varepsilon$ values. Such region is plotted in blue in \cref{fig1_reginstab_q0} a) for $\varepsilon = 4$. 
Further, the limit of stability given by $\lambda_1>0$ can be computed explicitly from the second order term of the Taylor expansion, \cref{caseepsil_all_eq:1}. It is given by
\begin{equation}
\zeta(\frac{v_0}{v^*},\varepsilon, \textbf{q}\to 0)=\zeta_0= \frac{3}{4} \frac{v_0}{v^*} \pm \frac{1}{4}\sqrt{\left(\frac{v_0}{v^*}\right)^2 - \left(1 +\varepsilon^2 \right)},
\label{caseepsil_eq:1}
\end{equation}
which is plotted in \cref{fig1_reginstab_q0}: by a red dashed line in panel a) for $\varepsilon=4$, and by continuous lines for several values of  $\varepsilon$ in panel b).

Increasing $\varepsilon$ leads to a shift of the LW instability region to higher values of the self-propulsion speed, as shown in \cref{fig1_reginstab_q0}. 
This result from the linear stability analysis is consistent with the  phase behavior of chiral active particles reported in previous works, showing that active rotation generically hinders motility-induced phase separation \cite{Liao2018,bickmann2020chiral,mani2022}. In our formalism, this is evidenced by the shift of the instability region to higher values of $\frac{v_0}{v^*}$, at increasing $\varepsilon$. According to our theory, one needs larger self-propulsion speed  to eventually destabilise the homogeneous state and  reach a condensate of chiral active particles. 

\subsubsection{Short-wavelength instabilities}\label{shortwavelengthinst}

So far, we have focused on LW instabilities that signal the onset of a phase separation. However, the formalism we have derived allows us to study instabilities happening at any wave vector $\textbf{q}$, meaning $\lambda_1(q)>0$. Actually, our analysis predicts a short-wavelength (SW) instability for $\varepsilon\neq0$ over a broad range of parameter values. A finite $\textbf{q}^* > 0$ indicates the growth of some structure, or pattern, with a characteristic length scale $\ell \sim 1/\textbf{q}^*$. Therefore, in the SW instability region of the parameter space, a phase separation (i.e. MIPS) is not expected, but rather the formation of smaller finite-sized clusters, which, according to the prediction to linear order, will not coarsen to form a macroscopic structure. 

To identify the  onset of a SW instability we perform an adiabatic approximation (i. e. $\partial_t \textbf{p} =0$) in \cref{modeldescrip_eq:16}, which allows us to rewrite the hydrodynamic equation for the density field as an effective diffusion equation (see \cref{alternative_adiabatic} for the full derivation). From it, one can obtain the  following limit of stability
\begin{equation}
\begin{split}
\zeta(\frac{v_0}{v^*},\varepsilon, \textbf{q})=\zeta_{\textbf{q}}= \frac{3}{4} \frac{v_0}{v^*} \pm \frac{1}{4}\sqrt{\left(\frac{v_0}{v^*}\right)^2 - \textbf{q}^2 - 1 - \frac{\varepsilon^2}{\textbf{q}^2 +1}},
\end{split}
\label{alternat_adiab_eq:9}
\end{equation}
now given as a function of $\frac{v_0}{v^*}$, $\varepsilon$ and $q $. In \cref{longwavelengthinst}, we have already obtained the limit of stability for the particular case $q\rightarrow 0$, \cref{caseepsil_eq:1}. However, \cref{alternat_adiab_eq:9} is more general: it accounts for the limit of stability at any finite value of $q$.

\begin{figure}[h]
\includegraphics[trim=100 50 0 190,width=\columnwidth]{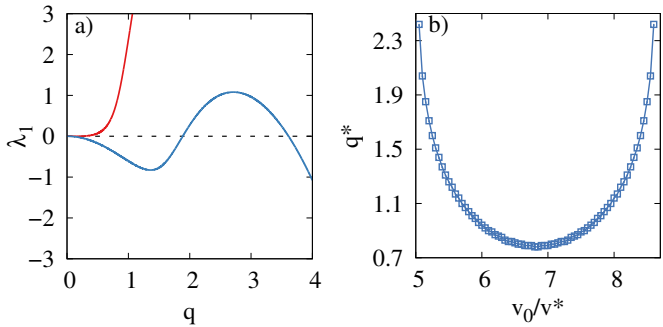}
   \caption{a) Eigenvalue responsible for the instability as a function of the wave vector for $(\frac{v_0}{v^*},\zeta,\varepsilon)=(8.0,6.0,8.0)$, corresponding to a long-wavelength instability (red line) and for $(\frac{v_0}{v^*},\zeta,\varepsilon)=(4.6,3.1,8.0)$, corresponding to a finite wavelength instability (blue line). b) Values of the wave vector $q^*$ at which the eigenvalue $\lambda_1$ first becomes positive as a function of $\frac{v_0}{v^*}$, for fixed $\zeta = 4.6$ and $\varepsilon = 8.0$ (horizontal dashed-dotted line in \cref{results_fig:2} b)). The solid line is a guide to the eye.}
    \label{results_fig:3} 
\end{figure}

We numerically compute the eigenvalues from the expressions in \cref{linear-stab-analys_eq:9}  and plot  $\lambda_1(q)$ for two representative cases in \cref{results_fig:3} a). In one case, $\lambda_1(q)$ is always positive irrespective of $q$ (red curve), while in the other case, it only becomes positive above a certain threshold $q^*$ (blue curve). The dependency of $q^*$ on $\frac{v_0}{v^*}$ at fixed $\varepsilon$ and $\zeta$ is shown in \cref{results_fig:3} b).

\begin{figure}[h]
\includegraphics[trim=300 20 0 0,width=\columnwidth]{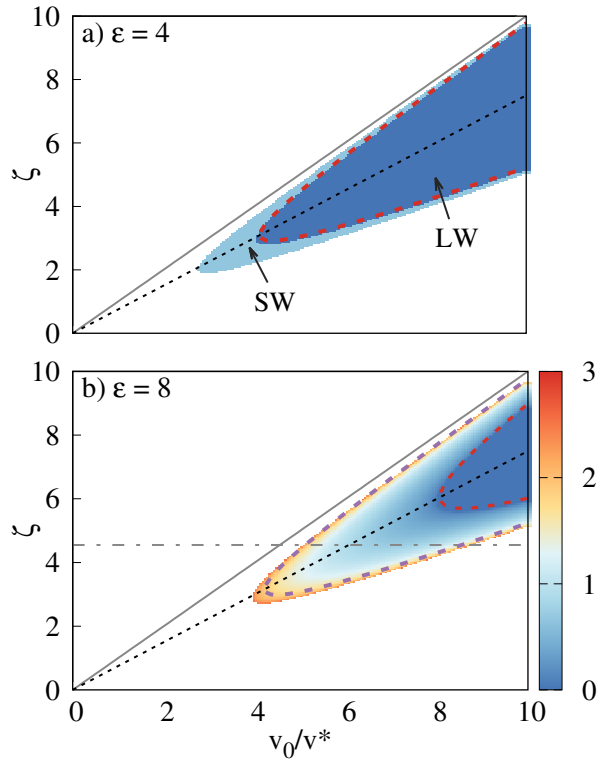}
   \caption{a) Region of instability at fixed $\varepsilon = 4$. The dark-blue region corresponds to the long-wavelength instability while the light blue region marks a region of short-wavelength instability. The red dashed curve indicates the limit of long-wavelength instability, predicted by \cref{caseepsil_eq:1}. b) Region of instability at fixed $\varepsilon= 8.0$. The color map corresponds to the value of $q^*$ at which the instability takes place. The dashed curves mark the limit of stability at fixed $q$, predicted by \cref{alternat_adiab_eq:9}. In this case, $q=0$ (red) and $q=1$ (purple), respectively. In both a) and b), the solid grey line corresponds to $v_{\bar{\rho}} = 0$ and the region above is nonphysical since $v_{\bar{\rho}} < 0$. The dotted black line indicates the line along which the critical point moves, upon increasing $\varepsilon$, given by \cref{evol_crit_point_eq:1} (\cref{evolcritpoint}).}
    \label{results_fig:2} 
\end{figure}

We observe that as soon as $\varepsilon \neq 0$ a SW instability appears, as depicted in \cref{results_fig:3} a). For $\varepsilon=4$, we plot in \cref{results_fig:2} a) the LW  instability region in dark blue together with the SW instability region in light blue. We also represent in \cref{results_fig:2} b) the SW and LW unstable regions for $\varepsilon=8$ with a color map  showing the value of $q^*$.  As evidenced  by the comparison between \cref{results_fig:2} a) and b), the extent of the SW instability region grows by increasing  $\varepsilon$. 
 Note that the closer a $(\frac{v_0}{v^*}, \zeta)$-point is to the long-wavelength (dark blue) instability region, the smaller the value of $q^*$ is, until it becomes identically zero inside of it. In other words, the characteristic length of the SW instabilities continuously grows when approaching the LW  instability region, until it becomes infinite (spanning all the system's size) inside it.

As we show in coloured dashed curves in \cref{results_fig:2} b), \cref{alternat_adiab_eq:9} successfully predicts both the LW and SW limits of stability. Thus, the dashed curves can be interpreted as 'iso-$q$lines' along which the instability will have the same characteristic length scale.  

Our analysis also allows for a quantification of the finite wavelength instability $q^*$ 
as a function of the self-propulsion speed $\frac{v_0}{v^*}$. We show such dependency, $q^*(\frac{v_0}{v^*})$, in \cref{results_fig:3} b) at fixed $\zeta= 4.6$, for which the system never enters the LW instability region (see also the dashed-dotted line in \cref{results_fig:2} b)). We report a decrease in $q^*$ as the system penetrates into the instability region, until it reaches a minimum and then monotonically grows before exiting towards the stable   region.

\begin{figure}
\centering
  \includegraphics[trim=300 50 150 210,width=0.5\columnwidth]{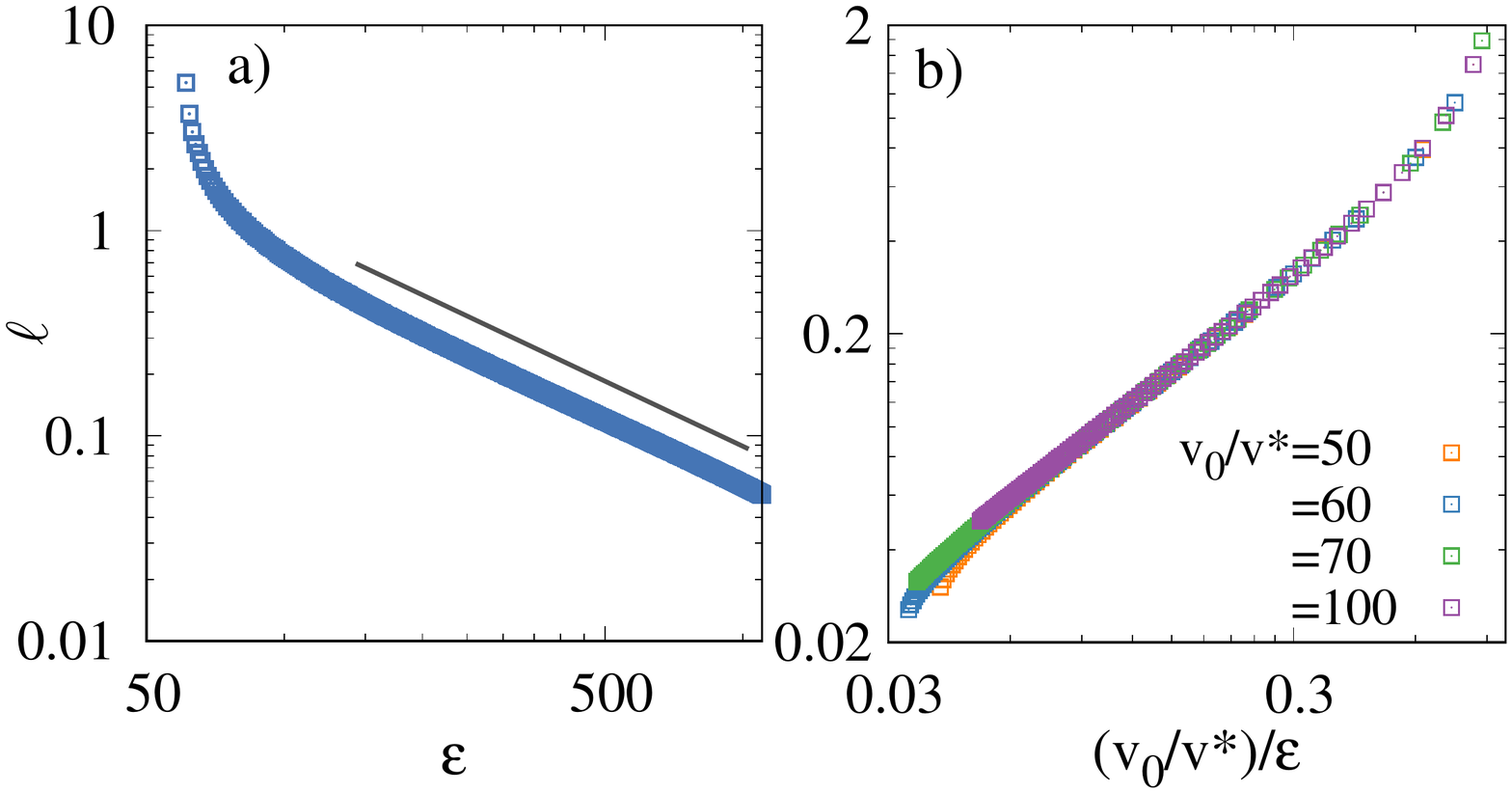}
  \caption{{a) Characteristic length scale of the short-wavelength instability, $\ell$, as a function of the rotational frequency $\varepsilon$ for a system at fixed $\zeta=45$ and $\frac{v_0}{v^*}=60$. b) Characteristic length scale as a function of the inverse rotational frequency normalized by the corresponding self-propulsion speed, $\frac{v_0}{v^*\varepsilon}$, for four different values of $(\frac{v_0}{v^*}, \zeta)_c = \{(50.0,37.5), (60.0,45.0), (70.0,52.5), (100.0,75.0)\}$. These fixed values correspond to points laying on the critical line (see \cref{evol_crit_point_eq:1} in \cref{evolcritpoint}).}
  }
  \label{results_fig:2qeps}
\end{figure}

In \cref{results_fig:2qeps} a), we look further at the dependency of  $\ell=1/q^*$ with $\varepsilon$. We now fix $\zeta =45$ and $\frac{v_0}{v^*}=60$,  allowing us to explore a broad range of  $\varepsilon$ values. We recall that the bigger $\varepsilon$ is, the larger the area of SW instability in the $(\frac{v_0}{v^*}, \zeta)$ plane. We find that $\ell$ decays as $1/\varepsilon$ over a broad range of parameter values. 
Consistently,  in the limit of $\varepsilon \rightarrow \infty $, the unstable eigenvalue $\lambda_1 \sim \varepsilon$ (see \cref{linear-stab-analys_eq:9}). Therefore, the limit of stability at finite wave vector, $\lambda_1(\frac{v_0}{v^*},\zeta,\varepsilon,q^*) = 0$, leads to $q^* \sim \varepsilon$ in the limit of large $\varepsilon$.  These results predict that the  rotational frequency of chiral active particles controls the selection of a characteristic length scale, which decreases with increasing $\varepsilon$. 
Further, in \cref{results_fig:2qeps} b), we plot $\ell$ as a function of the inverse rotational frequency for different fixed values of $\frac{v_0}{v^*}$ and $\zeta$, which all lay on the critical-point line defined in \cref{evolcritpoint},  \cref{evol_crit_point_eq:1}, and represented by a dotted line in \cref{results_fig:2} b). We observe a linear dependency $\ell \propto \frac{v_0}{v^*\varepsilon} $, indicating that the selected length scale is proportional to the radius of the individual circular trajectory of a chiral active particle, or circle swimmer.
Interestingly, a similar scaling has been found in suspensions of spinning magnetic rotors \cite{massana2021prr} and model systems of polar chiral active particles \cite{liebchen2017}. However, there are important differences between these systems and the present model, which are worth to be mentioned. First, magnetic colloids self-spin without net self-propulsion, while in our framework, no instability can take place in the limit $v_0\to 0$. Second,  in \cite{liebchen2017}, rotations trigger a SW instability of the homogenous polar (or flocking) state. A symmetry breaking has to occur in this case to give rise to microflocks of typical size $\ell \propto {v_0}/ \omega_0 $, while the continuum theory derived here does not admit solutions with global polar order. One has thus to be cautious when making connections between these different systems, although the fact that in all cases one finds a typical length scale $\ell\sim\omega^{-1}$, certainly deserves to be highlighted,  as it  suggests that  a common general pattern formation mechanism might be at play in systems of chiral particles.  

\section{Conclusions}
We have presented a general continuum description of self-propelled particles subjected to generic torques,  derived by explicitly coarse-graining the microscopic dynamics. As a consequence, the parameters of the hydrodynamic model are linked to the microscopic interactions as well as to the inter-particle spatial and angular correlations. Thus, we can interpret them based on the specific interactions that might come into play. This particular feature does not constraint the field equations but, on the contrary, it allows to describe a wide variety of particle-based models, where the force needs only be central, and torques can both be intrinsic to the particle (chiral) or derive from an alignment interaction of general functional form.

At the mean-field level, the linear stability analysis of the field equations unveils different instabilies of the homogeneous and isotropic state. We observe that torques tend to oppose to a long-wavelength instability, which we interpret as motility-induced phase separation. This is in agreement  with previous numerical studies \cite{Liao2018} as well as analytical descriptions of chiral active particles \cite{bickmann2020chiral,mani2022}. Moreover, effective torques lead to a finite wavelength instability, which suggests the formation of finite-sized structures. Our analysis predicts a linear dependency between the characteristic cluster size and the average radius of the trajectory of a single chiral active particle, 
$\ell\propto v_0/\varepsilon$. This result echoes the $\ell\sim\omega^{-1}_0$ behaviour  found in systems of self-spinning colloids and polar chiral active particles \cite{massana2021prr,liebchen2017}. 

Our analytical approach constitutes a powerful tool to investigate the effect that different interactions have on the destabilisation of MIPS. Besides, it is not constraint to a particular model but can be systematically applied to a number of dry active particle models by just fine-tuning their mutual interactions. Since the field theory is derived  from the microscopic dynamics, it also allows for a direct quantitative comparison with particle-based simulations.

An interesting continuation to this work would be to envision a particle-based model that breaks the angular symmetries $\varphi \rightarrow - \varphi$ and $\theta \rightarrow -\theta$ yielding $\kappa \neq 0$. This would allow to test the predictions given by the coarse-grained mean-field model in a system with effective aligning torques beyond chiral particles with a self-torque.

\section*{Conflicts of interest}
The authors have no conflicts to disclose.

\section*{Acknowledgements}
E.S.-S. and I.P. acknowledge the Swiss National Science Foundation Project No. 200021-175719. D.L. acknowledges MCIU/AEI/FEDER for financial support under Grant Agreement No. RTI2018-099032-J-I00. I.P. acknowledges support from Ministerio de Ciencia, Innovaci\'on y Universidades MCIU/AEI/FEDER for financial support under grant agreement PGC2018-098373-B-100 AEI/FEDER-EU and from Generalitat de Catalunya under project 2017SGR-884.

\newpage
\newpage

\appendix

 \section{Gram-Schmidt orthonormalization}\label{appx_gram_schmidt}

We would like to decompose the force in the vector basis $\{ \textbf{e}, \nabla \psi_1\}$. In order to ensure that this basis is orthonormal we perform a Gram-Schmidt orthonormalization. 

We pick the first vector of the orthonormal set $\{ \textbf{u}_1, \textbf{u}_2\}$ we want to construct, $\textbf{u}_1 = \textbf{e}$. This one already fulfils $|\textbf{e}| = 1$. Then the second vector fulfils $ \textbf{u}_2 = \nabla \psi_1 - \mathrm{proj}_{ \textbf{u}_1}\left( \nabla \psi_1 \right)$, which normalized leads to,

\begin{equation}
\begin{split} 
 \textbf{u}_2 = \frac{\nabla \psi_1 - \left( \textbf{e} \cdot \nabla \psi_1 \right) \textbf{e}}{|\nabla \psi_1 - \left( \textbf{e} \cdot \nabla \psi_1 \right) \textbf{e}|}.
\end{split}
\label{appx_gram_schmidt-eq:3}
\end{equation}
We have constructed an orthonormal vector basis. We can thus decompose the force as, 

\begin{equation}
\begin{split} 
 \textbf{F} &= \left( \textbf{e} \cdot  \textbf{F} \right)\textbf{e} + \left(  \frac{\nabla \psi_1 - \left( \textbf{e} \cdot \nabla \psi_1 \right) \textbf{e}}{|\nabla \psi_1 - \left( \textbf{e} \cdot \nabla \psi_1 \right) \textbf{e}|} \cdot   \textbf{F} \right) \frac{\nabla \psi_1 - \left( \textbf{e} \cdot \nabla \psi_1 \right) \textbf{e}}{|\nabla \psi_1 - \left( \textbf{e} \cdot \nabla \psi_1 \right) \textbf{e}|} \\
 &= \left[\left( \textbf{e} \cdot  \textbf{F} \right) - \left(  \frac{\left(\nabla \psi_1 - \left( \textbf{e} \cdot \nabla \psi_1 \right) \textbf{e} \right)\cdot   \textbf{F} }{|\nabla \psi_1 - \left( \textbf{e} \cdot \nabla \psi_1 \right) \textbf{e}|^2} \right) \left( \textbf{e} \cdot \nabla \psi_1 \right)\right]\textbf{e} \\
 &\qquad+ \left(  \frac{\left(\nabla \psi_1 - \left( \textbf{e} \cdot \nabla \psi_1 \right) \textbf{e} \right)\cdot   \textbf{F} }{|\nabla \psi_1 - \left( \textbf{e} \cdot \nabla \psi_1 \right) \textbf{e}|^2} \right)\nabla \psi_1 .
\end{split}
\label{appx_gram_schmidt-eq:4}
\end{equation}

Now we first consider that the projection of the force in the perpendicular vector to $\textbf{e}$ (this is, $\frac{\nabla \psi_1 - \left( \textbf{e} \cdot \nabla \psi_1 \right) \textbf{e}}{|\nabla \psi_1 - \left( \textbf{e} \cdot \nabla \psi_1 \right) \textbf{e}|}$) is much smaller than the projection of the force along $\textbf{e}$. Second, we also consider that $|\nabla \psi_1 - \left( \textbf{e} \cdot \nabla \psi_1 \right) \textbf{e}|^2 \approx |\nabla \psi_1 |^2$, assuming that $\textbf{e} $ and $\nabla \psi_1$ are 'almost' perpendicular vectors. This leads to, 

\begin{equation}
\begin{split} 
 \textbf{F} & \approx  \left( \textbf{e} \cdot  \textbf{F} \right) \textbf{e}+ \left(  \frac{\left(\nabla \psi_1 - \left( \textbf{e} \cdot \nabla \psi_1 \right) \textbf{e} \right)\cdot   \textbf{F} }{|\nabla \psi_1|^2} \right)\nabla \psi_1.
\end{split}
\label{appx_gram_schmidt-eq:5}
\end{equation}

\section{Eigenvalues' computation}\label{eigenval_compu}

To numerically compute the eigenvalues we express them in polar form, which makes it easier to deal with complex cubic roots. Below, we give a detailed explanation of how we compute them numerically. Solving the determinant of the matrix $M$ leads to a third degree polynomial of the form 

\begin{equation}
\lambda^3 + (a + 2b) \lambda^2 + (2 a b + c+ d) \lambda + c a + bd = 0,
\label{eigenval_compu_eq:1}
\end{equation}
where we have defined  the parameters a, b, c and d, corresponding to

\begin{equation}
\begin{split}
a&= \textbf{q}^2,\\
b&= \textbf{q}^2 +1,\\
c&=(\textbf{q}^2 +1)^2 +\varepsilon^2,  \\
d&=8 \textbf{q}^2(\frac{v_0}{v^*}-\zeta)(\frac{v_0}{v^*}-2\zeta) ,
\end{split}
\label{eigenval_compu_eq:2}
\end{equation}
in an attempt to make expressions shorter and notation clearer. 
The three solutions of the third degree polynomial, \cref{eigenval_compu_eq:1}, can be written as

 \begin{equation}
\begin{split}
\lambda_{i} = - \frac{a+2b}{3} - \frac{C_{i}}{3} - \frac{- A}{C_{i}},
\end{split}
\label{eigenval_compu_eq:3}
\end{equation}
where $i = 0,1,2$ and we define $A$, $B$ and $C_i$ as

\begin{equation}
\begin{split}
A=-(a+&2b)^2  +3(2 a b + c+d), \\
B=-2a^3 + 6 a^2b + 12ab^2 &-16b^3-18ac+18bc+9ad-9bd,
\end{split}
\label{eigenval_compu_eq:4}
\end{equation}

\begin{equation}
\begin{split}
C_{i} = - \sqrt[3]{\frac{B+\sqrt{B^2+4A^3}}{2}}.
\end{split}
\label{eigenval_compu_eq:5}
\end{equation}
The parameters $A$, $B$ and $C_{i}$, which in turn group combinations  of parameters $a$,  $b$,  $c$ and  $d$, have been introduced, again, for ease of notation. 

$C_{i}$ has three possible values ($i = 0,1,2$) corresponding to the three solutions of the cube root,

\begin{equation}
\begin{split}
C_0 =  - \sqrt[3]{\frac{B+\sqrt{B^2+4A^3}}{2}},            \\
C_1 = \left(-\frac{1}{2} + i \frac{1}{2} \sqrt{3}\right)C_0,   \\
C_2 = \left(-\frac{1}{2} - i \frac{1}{2} \sqrt{3}\right)C_0.      
\end{split}
\label{eigenval_compu_eq:6}
\end{equation}
Inserting the expression of $C_i$ for $i=0,1,2$ in \cref{eigenval_compu_eq:3} we obtain

\begin{equation}
\begin{split}
\lambda_{1}&=-\frac{1}{3}(a+2b) +\frac{1}{2^{1/3}3} (B-\sqrt{B^2 + 4 A^3})^{1/3}\\
&\qquad+ \frac{1}{2^{1/3}3}(B+\sqrt{B^2 + 4 A^3})^{1/3},\\
\lambda_{2}&=-\frac{1}{3}(a+2b)\\
&\qquad -\frac{1}{2^{4/3}3}\left[  (B+\sqrt{B^2 + 4 A^3})^{1/3} + (B-\sqrt{B^2 + 4 A^3})^{1/3} \right] \\
&\qquad +\frac{i}{\sqrt{3}\:2^{4/3}} \left[ (B+\sqrt{B^2 + 4 A^3})^{1/3} - (B-\sqrt{B^2 + 4 A^3})^{1/3} \right], \\
\lambda_{3}&=-\frac{1}{3}(a+2b) \\
&\qquad -\frac{1}{2^{4/3}3}\left[ (B+\sqrt{B^2 + 4 A^3})^{1/3} + (B-\sqrt{B^2 + 4 A^3})^{1/3} \right] \\
&\qquad -\frac{i}{\sqrt{3}\:2^{4/3}} \left[  (B+\sqrt{B^2 + 4 A^3})^{1/3}-(B-\sqrt{B^2 + 4 A^3})^{1/3} \right].
\end{split}
\label{linear-stab-analys_eq:9}
\end{equation}

\section{Linear stability analysis - alternative way: Adiabatic approximation} \label{alternative_adiabatic}

We start from the Fourier transformed effective hydrodynamic equations, which we write here for the purpose of clarity,

\begin{equation}
\begin{split}
\partial_{t} \delta \hat{ \rho}  =- i\textbf{q} \cdot \Big[ (v_0 - \bar{\rho}\zeta) \delta\hat{\textbf{p}} - \Da i \textbf{q}\delta\hat{\rho} \Big],
\end{split}
\label{alternat_adiab_eq:1}
\end{equation}

 \begin{equation}
\begin{split}
\partial_{t} \delta \hat{\textbf{p}} = - i \textbf{q} \cdot \big[  \frac{1}{2}(v_0 -2\bar{\rho} \zeta) \delta \hat{\rho} -\Da  i \textbf{q} \delta \hat{\textbf{p}} \big]  - \bar{\rho}\varepsilon  \delta\hat{\textbf{p}} ^{\perp}  - D_r \delta \hat{\textbf{p} }.
\end{split}
\label{alternat_adiab_eq:2}
\end{equation}
Note that we have not yet rewritten them in terms of dimensionless quantities. We now perform the adiabatic approximation by setting $\partial_{t} \delta \hat{\textbf{p}}=0$, which allows us to rewrite \cref{alternat_adiab_eq:2} as,

 \begin{equation}
\begin{split}
  i \textbf{q}  \frac{1}{2}(v_0 -2\bar{\rho} \zeta) \delta \hat{\rho} =  -  \textbf{q}^2\Da \delta \hat{\textbf{p}}   -  \bar{\rho}\varepsilon R \delta\hat{\textbf{p}}  -D_r \delta \hat{\textbf{p} },
\end{split}
\label{alternat_adiab_eq:3}
\end{equation}
where we have taken into account that $\bold{p}^{\perp}=\mathcal{R}\bold{p}$, with  $\mathcal{R} = \left(\begin{array}{ccc} 0 & -1\\1 & 0\\ \end{array}\right)$. We can thus express \cref{alternat_adiab_eq:3} as

 \begin{equation}
\begin{split}
  i \textbf{q}  \frac{1}{2}(v_0 -2\bar{\rho} \zeta) \delta \hat{\rho} =\textbf{A} \delta \hat{\textbf{p}},
\end{split}
\label{alternat_adiab_eq:4}
\end{equation}
where

 \[
 \textbf{A}=
  \begin{bmatrix}
   -  (\textbf{q}^2\Da + D_r)  &   \bar{\rho} \varepsilon \\
    - \bar{\rho} \varepsilon &-  (\textbf{q}^2\Da + D_r)
  \end{bmatrix}.
 \] 
It is now possible to compute the inverse matrix

 \[
 \textbf{A}^{-1}= \frac{1}{(\textbf{q}^2\Da + D_r) ^2 + (\bar{\rho} \varepsilon)^2}
  \begin{bmatrix}
   -  (\textbf{q}^2\Da + D_r)  &   -\bar{\rho} \varepsilon \\
    \bar{\rho} \varepsilon &-  (\textbf{q}^2\Da + D_r)
  \end{bmatrix},
 \] 
 allowing one to rewrite \cref{alternat_adiab_eq:4} as,
 
  \begin{equation}
\begin{split}
\delta \hat{\textbf{p}} = \textbf{A}^{-1} i \textbf{q}  \frac{1}{2}(v_0 -2\bar{\rho} \zeta) \delta \hat{\rho} .
\end{split}
\label{alternat_adiab_eq:5}
\end{equation}

From now on we will call $\textbf{C} = i \textbf{q}  \frac{1}{2}(v_0 -2\bar{\rho} \zeta)$ to shorten notation. We can now insert $\delta \hat{\textbf{p}} = \textbf{A}^{-1} \textbf{C}  \delta \hat{\rho} $ into the density equation, \cref{alternat_adiab_eq:1}, leading to,

\begin{equation}
\begin{split}
\partial_{t} \delta \hat{ \rho}  =  \Big[ -i  \textbf{q}(v_0 - \bar{\rho}\zeta) \textbf{A}^{-1} \textbf{C}- \Da \textbf{q}^2 \Big]\delta\hat{\rho} .
\end{split}
\label{alternat_adiab_eq:6}
\end{equation}
We have thus recasted  \cref{alternat_adiab_eq:1} into a diffusion equation, where the effective diffusion coefficient is the operator

\begin{equation}
\begin{split}
\mathcal{L} = -i  \textbf{q}(v_0 - \bar{\rho}\zeta) \textbf{A}^{-1} \textbf{C}- \Da \textbf{q}^2. 
\end{split}
\label{alternat_adiab_eq:7}
\end{equation}

The onset of destabilization of the homogeneous and isotropic solution can be identified when  $\mathcal{L}  < 0$. This leads to the closed expression,


\begin{equation}
\begin{split}
\zeta = \frac{3}{4} \frac{v_0}{\bar{\rho}} \pm  \frac{1}{4 \bar{\rho}}\sqrt{v_0^2 - 16 \Da^2 \textbf{q}^2 -16 \Da D_r  - 16 \Da \frac{(\bar{\rho}\varepsilon)^2}{\textbf{q}^2 +1}}.
\end{split}
\label{alternat_adiab_eq:8}
\end{equation}
Introducing the dimensionless quantities defined in Eq. (24), one can rewrite \cref{alternat_adiab_eq:8} as,
\begin{equation}
\begin{split}
\zeta= \frac{3}{4} \frac{v_0}{v^*} \pm \frac{1}{4}\sqrt{\left(\frac{v_0}{v^*}\right)^2 - \textbf{q}^2 - 1 - \frac{\varepsilon^2}{\textbf{q}^2 +1}}.
\end{split}
\label{alternat_adiab_limitstab_eq:9}
\end{equation}

\section{Evolution of the critical point}\label{evolcritpoint}

The vertex of the region of instability fulfils $\zeta^{+} = \zeta^{-}$, where $\zeta$ follows \cref{alternat_adiab_limitstab_eq:9}.
Thus, it is straightforward to find the location of the vertex for any value of $\textbf{q}$ and $\varepsilon$. 
We first apply $\zeta^{+} = \zeta^{-}$, to find $( \frac{v_0}{v^*})_{c} = \sqrt{\textbf{q}^2 + 1 + \frac{\varepsilon^2}{\textbf{q}^2 +1}}$. Inserting now this expression in \cref{alternat_adiab_limitstab_eq:9} we obtain the point in the instability region, which reads

\begin{equation}
\begin{split}
\left( \frac{v_0}{v^*},\zeta \right)_c = \left(\sqrt{ \textbf{q}^2 + 1 + \frac{\varepsilon^2}{\textbf{q}^2 +1}}, \frac{3}{4}\sqrt{ \textbf{q}^2 + 1 + \frac{\varepsilon^2}{\textbf{q}^2 +1}}\right),
\end{split}
\label{evol_crit_point_eq:1}
\end{equation}
as a function of $\textbf{q}$ and $\varepsilon$. Thus, varying $\textbf{q}$ and $\varepsilon$ the critical point moves along $\zeta =  \frac{3}{4} \frac{v_0}{v^*}$.

\newpage
\bibliography{coarse-grained_ABP_weak_alignment}
\end{document}